%% file: paper.tex
\documentclass{sig-alternate}

\usepackage{times}
\usepackage{url}
\usepackage{subfigure}
\usepackage{color}
\usepackage{ifthen}
\usepackage{algorithmic}
\usepackage{algorithm}
\usepackage{graphicx}

\newcommand{\comment}[1]{}
\newcommand{\tpushend}{\ensuremath{{\mathrm{push\_end}}}}
\newcommand{\tmapstart}{\ensuremath{{\mathrm{map\_start}}}}
\newcommand{\tmapend}{\ensuremath{{\mathrm{map\_end}}}}
\newcommand{\tshufflestart}{\ensuremath{{\mathrm{shuffle\_start}}}}
\newcommand{\tshuffleend}{\ensuremath{{\mathrm{shuffle\_end}}}}
\newcommand{\treducestart}{\ensuremath{{\mathrm{reduce\_start}}}}
\newcommand{\treduceend}{\ensuremath{{\mathrm{reduce\_end}}}}

\newcommand{\unit}[1]{\ensuremath{\, \mathrm{#1}}}

\newenvironment{myitemize}
{
  \begin{list}{- \ }{}
    \setlength{\topsep}{0pt}
    \setlength{\parskip}{0pt}
    \setlength{\partopsep}{0pt}
    \setlength{\parsep}{0pt}         
    \setlength{\itemsep}{0pt} 
}
{
  \end{list} 
}

\makeatletter
\let\@copyrightspace\relax
\makeatother

\begin{document}

\title{Optimizing MapReduce for Highly Distributed Environments\thanks{This work was supported in part by NSF Grants CNS-0643505 and CNS-0519894.}}

\numberofauthors{3}
\author{
  \alignauthor Benjamin Heintz\\
    \affaddr{University of Minnesota}\\
    \affaddr{Minneapolis, MN}\\
    \email{heintz@cs.umn.edu}
  \alignauthor Abhishek Chandra\\
    \affaddr{University of Minnesota}\\
    \affaddr{Minneapolis, MN}\\
    \email{chandra@cs.umn.edu}
  \alignauthor Ramesh K. Sitaraman\\
    \affaddr{University of Massachusetts}\\
    \affaddr{Amherst, MA}\\
    \email{ramesh@cs.umass.edu}
}

\maketitle
\thispagestyle{empty}

\input{Abstract}
\input{Intro}
\input{Model}
\input{Implementation}
\input{OptEval}
\input{ModelPreds}
\input{Related}
\input{Conc}

\bibliographystyle{abbrv}
%\bibliography{refs}

\end{document}

%% file: Abstract.tex
\begin{abstract}

MapReduce, the popular programming paradigm for large-scale data processing,
has traditionally been deployed over tightly-coupled clusters where the data is
already locally available.
The assumption that the data and compute resources are available in a single
central location, however, no longer holds for many emerging applications in
commercial, scientific and social networking domains, where the data is
generated in a geographically distributed manner.
Further, the computational resources needed for carrying out the data analysis
may be distributed across multiple data centers or community resources such as
Grids.
In this paper, we develop a modeling framework to capture MapReduce execution
in a highly distributed environment comprising distributed data sources and
distributed computational resources.
This framework is flexible enough to capture several design choices and
performance optimizations for MapReduce execution.
We propose a model-driven optimization that has two key features: (i) it is
end-to-end as opposed to myopic optimizations that may only make locally
optimal but globally suboptimal decisions, and (ii) it can control multiple
MapReduce phases to achieve low runtime, as opposed to single-phase
optimizations that may control only individual phases.
Our model results show that our optimization can provide nearly 82\% and 64\%
reduction in execution time over myopic and single-phase optimizations,
respectively.
We have modified Hadoop to implement our model outputs, and using three
different MapReduce applications over an 8-node emulated PlanetLab testbed, we
show that our optimized Hadoop execution plan achieves 31-41\% reduction in
runtime over a vanilla Hadoop execution.
Our model-driven optimization also provides several insights into the choice of
techniques and execution parameters based on application and platform
characteristics.

\end{abstract}

%% file: Intro.tex
\section{Introduction}
\label{sec:intro}

\subsection{Motivation}

The growing need for analysis of large quantities of data generated globally by
increasing numbers of users, applications, sensors, and devices has led to wide
popularity of the MapReduce~\cite{mapreduce} model and its open-source
Hadoop~\cite{hadoop} implementation.
MapReduce is widely used today for many data analysis applications, including
for example Web indexing, log file analysis, and recommendation mining.

MapReduce has traditionally been deployed over a tightly-coupled cluster or
data center, with the assumption that the data is already available at a
centralized location, co-located with the computational resources.
For many emerging applications and environments, however, data sources are
inherently distributed, and the assumption of centralized data and centralized
computational resources does not hold.
For instance, a number of technology companies as well as traditional
businesses such as retail~\cite{retail-usecase} generate data at multiple sites
including stores and warehouses situated in diverse locations.
Further, large Internet-scale services such as Akamai \cite{NygrenSS10} have
highly distributed server deployments spanning thousands of global locations.
Each location produces tens of terabytes of data daily, and this data must be
aggregated and analyzed.
As further examples, many data-intensive scientific applications need to
analyze data generated by distributed scientific sensors, instruments, and
testbeds, while the data for several social networking applications is
generated by millions of users around the world.
For such applications, moving data to a central location for analysis can be
extremely costly, both in time and money~\cite{berkeleycloud}.

Besides the data sources being distributed, the computational resources needed
for carrying out the data analysis may themselves be distributed.
Examples include multiple data centers that may be used by a large Internet
company to analyze data local to these data centers, as well as community
resources such as Grids used for scientific computation.
The ever-growing need for efficient execution of MapReduce computations in
highly-distributed environments is the key motivator of our work.

MapReduce efficiency is itself a well-studied problem.
Several
techniques~\cite{heteromapred,ananthanarayanan:osdi2010,condie:nsdi2010} have
been proposed to improve the performance of MapReduce in local cluster
environments.
It is unclear, however, which of these techniques (if any) are well-suited for
executing MapReduce in highly distributed environments.
Recent work~\cite{dmr11} that explored deploying MapReduce in a highly
distributed environment concluded that there is no single architecture or
deployment strategy that works well for all possible application, data,
network, and system characteristics.
Thus, the tradeoffs of deploying and executing MapReduce in a highly
distributed environment are not well understood.
Further, few guidelines exist on how to efficiently execute a MapReduce
application in such an environment.
The focus of our work is to provide such guidelines by modeling the
characteristics of the application and the distributed environment and devising
an optimized ``execution plan'' that can guide the efficient execution of a
MapReduce job.

\subsection{The MapReduce framework}
\begin{figure}[htb]
  \begin{center}
  \epsfig{file=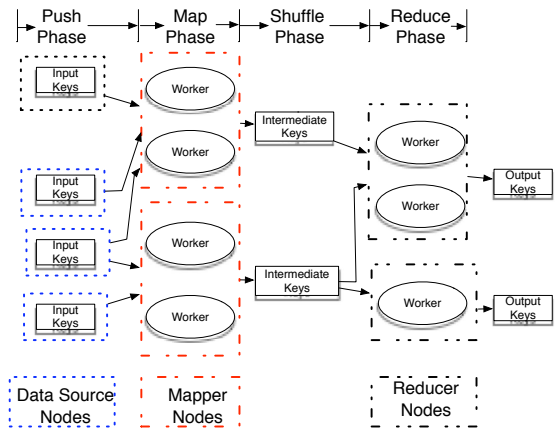,width=3in}
  \caption{Executing a MapReduce application in a distributed environment.
  Each node represents a cluster of machines deployed in a data center.
  In a highly distributed and heterogeneous environment, the nodes have varying
  amounts of resources distributed across a wide-area network.}
  \label{fig:mapreduce}
  \end{center}
\end{figure}

The MapReduce programming framework can be used to implement a number of
commonly-used applications that take a set of {\em input} key/value pairs and
produce a set of {\em output} key/value pairs \cite{mapreduce}.
A typical MapReduce application consists of  a {\it map} function that
processes input key/value pairs from the input data to generate a set of {\it
intermediate} key/value pairs, and a {\it reduce} function that merges all
intermediate values associated with the same intermediate key to produce the
output key/value pairs.
We illustrate the execution of a typical MapReduce application in a highly
distributed environment in Figure~\ref{fig:mapreduce}.
The input data originates at the {\it data source nodes} and is distributed to
the {\it mapper nodes} in the {\it push phase}.
In the {\it map phase}, each mapper node that consists of multiple worker
threads running on multiple machines within a cluster performs the map
operation on the incoming data and outputs intermediate key/value pairs.
In the {\it shuffle phase}, the intermediate key/value pairs are partitioned
and distributed to the {\it reducer nodes} such that {\it all} records that
correspond to a given intermediate key are sent to the same reducer node.
This requirement preserves the semantics of a MapReduce application where {\it
all} values of a specific intermediate key are required to perform the correct
reduction.
In the {\it reduce} phase, the reducer nodes process the intermediate key/value
pairs and produce the final output.

Our notion of efficiency for the execution of a MapReduce job is {\em
makespan}, defined to be the total time taken for the job to complete.
Given a highly distributed platform in the form of multiple machine clusters
deployed in a wide-area network and given a MapReduce application, {\em our
goal is  to  optimize the makespan of executing the MapReduce application on
the distributed platform}.

\begin{figure}[htb]
  \begin{center}
  \epsfig{file=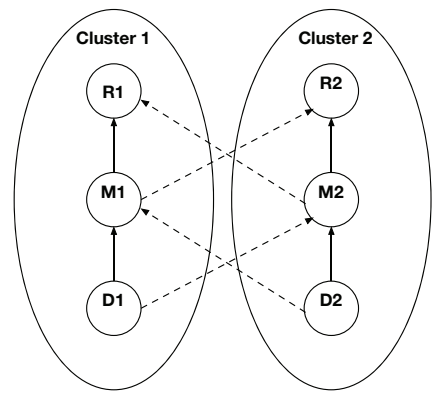, height=2in}
  \caption{An example of a two-cluster distributed environment, with one data
  source, one mapper, and one reducer in each cluster.
  The intra-cluster ``local'' communication links are solid lines, while the
  inter-cluster ``non-local'' communication links are the dotted lines.}
  \label{fig:mapreduceexample}
  \end{center}
\end{figure}

\subsection{An optimization example}

In order to optimize the makespan of MapReduce, we illustrate the criticality
of optimizing the data communication and task placement in a distributed
environment with a simple example, and show that the best approach depends on
the platform and application characteristics.

Consider the example MapReduce platform shown in
Figure~\ref{fig:mapreduceexample}.
Assume that the data sources $D1$ and $D2$ have 150\unit{GB} and 50\unit{GB} of
data, respectively.
Let us model a parameter $\alpha$ that represents the ratio of the amount of
data output by a mapper to its input; i.e., $\alpha$ is the expansion (or
reduction) of the data in the map phase and is specific to the application.
We model different situations below.

First, consider a situation where $\alpha = 1$ and the network is perfectly
homogeneous.
That is, all links have the same bandwidth---say 100\unit{MBps} each---whether
they are local or non-local, and the computational resources at each mapper and
reducer are homogeneous too; say, each can process data at the rate of
100\unit{MBps}.
Clearly, in this case, a uniform data placement, where each data source (resp.,
mapper) pushes an equal amount of data to each mapper (resp., reducer),
produces the minimum makespan.

Now, consider a slight modification to the above scenario.
Assume now that the non-local links become much slower and can only transmit at
10\unit{MBps}, while all other parameters remain the same.
Now, a uniform data placement no longer produces the best makespan.
Since the non-local links are much slower, it makes sense to avoid non-local
links when possible.
In fact, a plan where each data source pushes all of its data to its own local
mapper completes the push phase in $150 \unit{GB}/100 \unit{MBps} = 1500$
seconds, while the uniform push would have taken $75 \unit{GB}/10 \unit{MBps} =
7500$ seconds.
Although the map phase for the uniform placement would be smaller by $50
\unit{GB}/100 \unit{MBps} = 500$ seconds, the local push approach makes up for
this through a more efficient data push.

Finally, consider the same network parameters above where local links are fast
(100\unit{MBps}) and non-local links are slow (10\unit{MBps}).
However imagine that $\alpha$ is very large and equals 10, implying that there
is 10 times more data in the shuffle phase than in the push phase.
The local push no longer performs well, since it does nothing to alleviate the
communication bottleneck in the shuffle phase.
To avoid non-local communication in the shuffle phase, it makes sense for data
source $D2$ to push all of its 50\unit{GB} of data to mapper $M1$ in the push
phase, so that all the reduce can happen within cluster 1 without having to use
non-local links in the communication-heavy shuffle phase.
This is an example of how an optimization would have to look at {\em all}
phases in an {\em end-to-end} fashion to find a better makespan.
In fact, the local push still minimizes the push time from a {\em myopic}
perspective; i.e., it makes a locally optimal decision which is globally
suboptimal in terms of makespan.

The above simple example illustrates the types of considerations in speeding up
MapReduce in distributed environments.
Our goal in the rest of the paper is to model the distributed environment and
the MapReduce application accurately and to evolve techniques that can
automatically produce an optimal plan for executing the MapReduce job so as to
minimize makespan.

We note that there are two types of optimizations that can be used to speed up
a MapReduce computation: dynamic and static.
Dynamic optimization works at runtime as the MapReduce computation is being
executed.
Hadoop and other MapReduce software frameworks implement several dynamic
optimizations such as speculative task execution and work stealing.
In contrast to dynamic optimizations, static optimizations are performed even
before the MapReduce job starts execution, and are hence a complementary form
of optimization.
Our focus here is on static optimization and we are concerned with producing
the optimum data placement and task execution plan for a MapReduce job, taking
into account the underlying distributed data sources, network resources,
compute resources, and application characteristics.

\subsection{Research contributions}
Our paper makes the following research contributions:

\begin{myitemize}
  \item
  We develop a framework for modeling the performance of MapReduce in a highly
  distributed setting.
  Our modeling framework is flexible enough to capture a large number of design
  choices and optimizations.
  Our work provides a framework for answering ``what-if'' questions on the
  relative efficacy of various design alternatives.
  In particular, our model enables us to compare various architectural choices
  as well as to provide rules of thumb for efficient deployment based on
  network and application characteristics.
  Further, optimizations using our model are efficiently solvable as Mixed
  Integer Programs (MIP) using powerful solver libraries.

  \item
  We modify Hadoop to implement our model outputs and use this modified
  Hadoop implementation along with network and node measurements from the
  PlanetLab~\cite{planetlab} testbed to validate our model and evaluate our
  proposed optimization.
  Our model results show that for a highly distributed compute environment, our
  end-to-end, multi-phase optimization can provide nearly 82\% and 64\%
  reduction in execution time over myopic and the best single-phase
  optimizations, respectively.
  Further, using three different applications over an 8-node testbed emulating
  PlanetLab network characteristics, we show that our statically-enforced
  optimized Hadoop execution plan achieves 31-41\% reduction in runtime over a
  vanilla Hadoop execution employing its dynamic scheduling techniques.

  \item
  Our model-driven optimization provides several insights into the choice of
  techniques and execution parameters based on application and platform
  characteristics.
  For instance, we find that an application's data expansion factor $\alpha$
  can influence the optimal execution plan significantly, both in terms of
  which phases of execution are impacted more and where pipelining is more
  helpful.
  Our results also show that as the network becomes more distributed and
  heterogeneous, our optimization provides higher benefits (82\% for globally
  distributed sites vs. 37\% for a single local cluster over myopic
  optimization), as it can exploit heterogeneity opportunities better.
  Further, these benefits are obtained independent of the application
  characteristics.

\end{myitemize}

%% file: Model.tex
\section{Models and optimization algorithms}
\label{sec:model}
We successively model the distributed platform, the MapReduce application,
valid execution plans, and their makespan. 

\subsection{Modeling the distributed platform and the MapReduce application}
\begin{figure}[htbp]
  \centering
  \includegraphics[width=0.75\columnwidth]{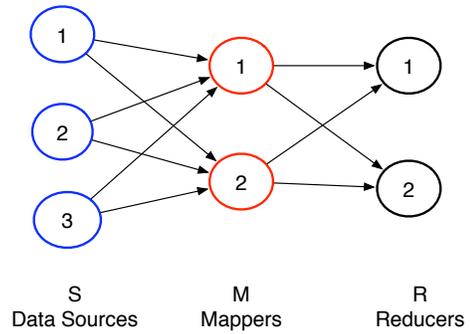}
  \caption{A tripartite graph model for distributed MapReduce with 3 data
  sources, 2 mappers and 2 reducers.\label{fig:tripartite_graph}}
\end{figure}

We model the distributed platform available for executing the MapReduce
application as a tripartite graph with a vertex set of $V = S \cup M \cup R$,
where $S$ is the set of data sources, $M$ is the set of mappers, and $R$ is the
set of reducers.
The edge set $E$ of the tripartite graph is the complete set of edges and
equals $(S \times M) \cup (M \times R)$ (See
Figure~\ref{fig:tripartite_graph}).
Each node corresponds to a cluster of servers that can potentially be used for
executing the MapReduce application and the edges represent communication paths
between clusters.
The {\em capacity of a node} $i \in M \cup R$ is denoted by $C_i$ (in
bits/second), where the capacity captures the computational resources available
at that node in the units of bits of incoming data that it can process per
second.
Note that $C_i$ is also application-dependent as different MapReduce
applications are likely to require different amounts of computing resources to
process the same amount of data.
Likewise, the {\em capacity of an edge} $(i, j) \in E$ is denoted by $B_{ij}$
that represents the bandwidth (in bits/second) that can be sustained on the
communication link that the edge represents. 

Rather than model the MapReduce application in detail, we model two key
parameters.
First, we model the {\em amount of data} $D_i$ (in bits) that originates at
data source $i$, for each $i \in S$.
Further, we model an {\em expansion factor} $\alpha$ that represents the ratio
of the size of the output of a mapper to the size of its input.
Note that $\alpha$ can take values less than, greater than, or equal to $1$
depending on whether the output of the map operation is smaller than, larger
than, or equal in size to the input.
The value of $\alpha$ can be determined by profiling the MapReduce application.
Many applications perform extensive aggregation, for example to count the
number of occurrences of each word in a set of documents, or to find the $k$
most extreme values in a large set.
These applications have $\alpha$ much less than 1.
On the other hand, some applications may require intermediate data to be
broadcast from one mapper to multiple reducers~\cite{rao2012}, or to otherwise
transform the inputs to produce larger intermediate data, yielding $\alpha>1$.

\subsection{Modeling a valid execution plan and its makespan}

We define the notion of an {\it execution plan} to capture the manner in which
data and computation of a MapReduce application are scheduled on a distributed
platform.
Intuitively, an execution plan determines how the data is distributed from the
sources to the mappers and how the intermediate keys produced by the mappers
are distributed across the reducers.
Thus, the execution plan determines which nodes and which communication links
are used and to what degree, and is therefore a major determinant of how long
the MapReduce application takes to complete. 

An execution plan of the MapReduce application on the distributed platform is
represented by variables $x_{ij}$, for $(i,j) \in E$, that represent the
fraction of the outgoing data from node $i$ that is sent to node $j$.
From a high level, an execution plan can be implemented by extending the
MapReduce software framework as follows.
Note that existing MapReduce frameworks such as Hadoop partition the input key
space at the data sources and assign each partition to a mapper.
Likewise, implementations partition the intermediate key space output by the
mappers, and assign each partition to a reducer.
As we mention later, this partitioning is usually achieved by a simple uniform
hash function that divides the key space evenly into the required number of
buckets.
Generalizing this concept, we can allow each data source and mapper to use its
own hash function to partition the input and intermediate key spaces
respectively, even perhaps in a non-uniform fashion.
This enables the implementation of any execution plan $x_{ij}$, for $(i,j) \in
E$, by simply providing a hash function $h_i$ to each node $i \in S$ (resp. $i
\in M$) such that a fraction $x_{ij}$ of the input keys (resp., intermediate
keys) are sent to mapper $j$ (resp., reducer $j$).
We describe such an approach more concretely in Section~\ref{subsec:impl}.

\noindent{\bf Valid execution plans.} 
We now mathematically express sufficient conditions for an execution plan to be
valid and implementable in a MapReduce software framework while obeying the
MapReduce application semantics.
\begin{eqnarray}
  \forall (i,j) \in E: 0 \leq x_{ij} \leq 1 \label{eq:frac1}\\
  \forall i \in V : \sum_{(i,j) \in E} x_{ij} = 1 \label{eq:frac2}
\end{eqnarray}
Equations~\ref{eq:frac1} and \ref{eq:frac2} simply express that for each $i$,
the $x_{ij}$'s are fractions that sum to $1$.

The semantics of a MapReduce application requires that each intermediate key be
mapped to a single reducer.
This is typically achieved by partitioning the intermediate key space among all
the reducers and ensuring that each reducer gets all the keys in its assigned
key space.
This semantics can be implemented by ensuring that all mappers use the {\it
same} hash function to map intermediate keys to reducers.
Define variable $y_k$, $k \in R$, to be the fraction of the key space mapped to
reducer $k$.\footnote{We assume that the key space is large enough so that the
variables $y_k$ can be accurately approximated.}
We enforce the one-reducer-per-key requirement with the following constraint.
\begin{equation}
  \forall j \in M, k \in R:  x_{jk} = y_k . \label{eq:onekeyrule}
\end{equation}
We define an execution plan to be {\it valid} if Equations~\ref{eq:frac1},
\ref{eq:frac2}, and \ref{eq:onekeyrule} hold.

\noindent{\bf Modeling the consecutive execution of phases.} 
The push, map, shuffle, and reduce phases are executed in sequence.
Between each pair of consecutive phases, there are three possible assumptions
that we can make.
The simplest assumption is that a {\em global barrier} exists between the two
consecutive phases.
That is, all nodes must complete any given phase before any node can proceed to
the next phase.
While the global barrier satisfies data dependencies across the different
phases, it has no concurrency in terms of the execution of the phases.
Alternately, one can assume that a {\em local barrier} exists between the two
consecutive phases.
With a local barrier, each node can start the next phase as soon as it receives
all its inputs from the previous phase {\em without} waiting for other nodes to
complete the previous phase.
For instance, with a local barrier, a node can start the map or reduce phase as
long as it has {\it all} of its input data, and a node can start the push and
shuffle phases as long as it has {\it all} the data that it needs to
disseminate.
In particular, a node need not wait for other nodes to complete the current
phase before proceeding to the next phase.
This allows a greater degree of concurrency between the execution of the
different phases, allowing the makespan to be reduced.
The third option of {\em pipelining} provides the greatest amount of
concurrency and has the potential for the lowest makespan among the three
options.
Pipelining allows a node to start the map or reduce phase as long as the {\it
first piece} of its input data is available for processing; i.e., the node need
not wait for all of its inputs to be present but rather it can start as soon as
it receives the first piece.
Likewise, a node can start the push and shuffle phases as soon as it receives
the {\it first piece} of the data that needs to be disseminated.
It is easy to see that pipelining allows for even more concurrency than
local barriers.

Which of these three options is allowable for each pair of consecutive phases
depends on application semantics and how the application itself is implemented.
For instance, for a specific application, a reducer may be able to proceed
independently of other reducers, but may need to wait until it receives all of
its intermediate data before it can start execution.
In this case, a local barrier is allowable, but pipelining is not.
Or, perhaps the application is such that incremental processing is possible in
each phase without receiving all of the input data, which is amenable to
pipelining.
We now show how to model each of these three possibilities starting with the
simplest case where all barriers are global.

\noindent{\bf Makespan of a valid execution plan with global barriers.} Given a
valid execution plan for a MapReduce application, we now model the total time
to completion, i.e., its makespan.
To model the makespan, we successively model the  time to completion of each of
the four phases assuming that a global barrier exists after each phase.
We assume that the data is available at all the data sources at time zero when
the push phase begins.
For each mapper $j \in M$, the time for the push phase to end is denoted by
$\tpushend_{j}$ that equals the time when all its data is received; i.e.,
\begin{equation}
  \forall j \in M : \tpushend_{j} =
    \max_{i \in S} \frac{D_{i} x_{ij}}{B_{ij}} \label{eq:pushend} .
\end{equation}
Since we assume a global barrier after the push phase, all mappers must receive
their data before the push phase ends and the map phase begins.
Thus, the time when the map phase starts denoted by $\tmapstart$ obeys the
following equation.
\begin{equation}
  \tmapstart = \max_{j \in M} \tpushend_{j} \label{eq:mapstart}
\end{equation}
The computation at each mapper takes time proportional to the data pushed to
that mapper.
Thus, the time $\tmapend_{j}$ for mapper $j \in M$ to complete its computation
obeys the following.
\begin{equation}
  \forall j \in M : \tmapend_{j} = \tmapstart +
    \frac{\sum_{i \in S} D_{i} x_{ij}}{C_{j}} \label{eq:mapend}
\end{equation}
As a result of the global barrier, the shuffle phase begins when all mappers
have finished their respective computations.
Thus, the time $\tshufflestart$ when the shuffle phase starts obeys the
following. 
\begin{equation}
  \tshufflestart = \max_{j  \in M} \tmapend_j \label{eq:shufflestart}
\end{equation}
The shuffle time for each reducer is governed by the slowest shuffle link into
that reducer.
Thus the time when shuffle ends at reducer $k \in R$ denoted by
$\tshuffleend_{k}$ obeys the following.
\begin{align}
  \lefteqn{\forall k \in R : \tshuffleend_{k} = } \nonumber \\ 
  & & \tshufflestart + \max_{j \in M} \frac{\alpha \sum_{i \in S} D_{i} x_{ij} x_{jk}}{B_{jk}}
  \label{eq:shuffleend}
\end{align}
Following the global barrier assumption, the reduce phase computation begins
only after all reducers have received all of their data.
Let $\treducestart$ be the time when the reduce phase starts.
Then
\begin{equation}
  \treducestart = \max_{k \in R} \tshuffleend_{k} . \label{eq:reducestart}
\end{equation}

Reduce computation at a given node takes time proportional to the amount of
data shuffled to that node.
Thus, the time when reduce ends at node $k$ denoted by $\treduceend_{k}$ obeys
the following.
\begin{align}
  \lefteqn{\forall k \in R : \treduceend_{k} = } \nonumber \\
  & & \treducestart + \frac{\alpha \sum_{i \in S} \sum_{j \in M} D_{i} x_{ij} x_{jk}}{C_{k}}
  \label{eq:reduceend}  
\end{align}

Finally, it is clear that the makespan equals the time at which the last
reducer finishes.
Thus
\begin{equation}
  \mathrm{Makespan} = \max_{k \in R} \treduceend_{k} \label{eq:makespan} .
\end{equation}

\noindent{\bf Modeling local barriers and pipelining.} We now show how to
modify the above constraints to model local barriers and pipelining.
First, we replace the constraints governing the start time for the last three
phases as expressed in Equations~\ref{eq:mapstart}, \ref{eq:shufflestart} and
\ref{eq:reducestart} with the following new constraints.
These new constraints capture the fact that a node can start its next phase
without waiting for all nodes to complete the previous phase.
\begin{equation*}
  \begin{split}
    \forall j \in M: \tmapstart_{j} &= \tpushend_{j} \\
    \forall j \in M: \tshufflestart_{j} &= \tmapend_{j} \\
    \forall k \in R: \treducestart_{k} &= \tshuffleend_{k}
  \end{split}
\end{equation*}

Now we can generalize the definitions for the ending times of these stages by
first defining a combination operator $\oplus$ for each optimization as follows:
\begin{equation*}
  a \oplus b = 
  \begin{cases}
    a + b, & \text{if local barrier} \\
    \max(a, b), & \text{if pipelined}
  \end{cases}
\end{equation*}

Then, we replace the definition for the ending time of the map phase in
Equation~\ref{eq:mapend} with the following new constraint.
\begin{equation}
  \forall j \in M : \tmapend_{j} =
  \tmapstart_{j} \oplus \frac{\sum_{i \in S} D_{i} x_{ij}}{C_{j}}
  \label{eq:map-end-general}
\end{equation}
Intuitively, the above constraint specifies that the runtime for the map phase
on a node depends on the start-time of the map phase on that node and the time
for map computation on all of the data pushed to that node.
For the local barrier case, this equals the sum of the two times (corresponding
to the time to push the data and then compute on the data in sequence).
On the other hand, for the pipelining case, this equals the maximum of the two
times, since the map phase at a node cannot end until all of its data has
arrived and all of its computation has finished, and the slower of these two
operations will dictate the completion time.
Note that we are assuming that the data push and map computation are completely
overlapped.
This assumption is valid if the total quantity of data is much larger than
individual record size, which is the case for many typical MapReduce
applications.

Based on similar intuition, the constraint for the shuffle stage in
Equation~\ref{eq:shuffleend} is replaced with
\begin{align}
  \lefteqn{\forall k \in R : \tshuffleend_{k} = } \nonumber \\
  & & \max_{j \in M} \left\{ \tshufflestart_{j} \oplus \frac{\alpha \sum_{i \in S} D_{i} x_{ij} x_{jk}}{B_{jk}}  \right\}
  \label{eq:shuffle-end-general}
\end{align}

Finally, the constraint for the end of the reduce stage in
Equation~\ref{eq:reduceend} is similarly replaced with
\begin{align}
  \lefteqn{\forall k \in R : \treduceend_{k} = } \nonumber \\
  & & \treducestart_{k} \oplus \frac{\alpha \sum_{i \in S} \sum_{j \in M} D_{i} x_{ij} x_{jk}}{C_{k}}
\end{align}

\subsection{Our optimization algorithm}
\label{subsec:optimize}
We formulate the problem of finding the execution plan that minimizes the
makespan as an optimization problem.
Viewing Equations~\ref{eq:frac1} to \ref{eq:makespan} as constraints, we need
to minimize a profit function that equals the variable $\mathrm{Makespan}$.
(Note that if the barriers are not global, the appropriate substitutions of the
constraints need to be made.)
To perform this optimization efficiently, we rewrite the constraints in linear
form to obtain a Mixed Integer Program (MIP).
Writing it as MIP opens up the possibility of using powerful off-the-shelf
solvers such as the Gurobi version 5.0.0 that we use to derive our results.

There are two types of nonlinearities that occur in our constraints, and these
need to be converted to linear form.
The first type is the existence of the max operator in
Equations~\ref{eq:pushend}, \ref{eq:mapstart}, \ref{eq:shufflestart},
\ref{eq:shuffleend}, \ref{eq:reducestart}, and \ref{eq:makespan}.
Here we use a standard technique of converting an expression of the form
$\max_i z_i = Z$ into an equivalent  set of linear constraints $\forall i: z_i
\leq Z$, where $Z$ is minimized as a part of the overall optimization. 

The second type of non-linearity arises from the quadratic terms in
Equations~\ref{eq:shuffleend} and \ref{eq:reduceend}.
We remove this type of nonlinearity with a two step process.
First, we express the product terms of the form $yz$ that involve two different
variables in separable form.
Specifically, we introduce two new variables $w = \frac{1}{2} \left( y + z
\right)$ and $w' = \frac{1}{2} \left( y - z \right)$.
This allows us to replace each occurrence of $yz$ with $w^2 - w'^2$, which is in
separable form since it does not involve a product of two different variables.
Next, we approximate the quadratic terms $w^2$ and $-w'^2$ with a piecewise
linear approximation of the respective function.
The more piecewise segments we use the more accuracy we achieve, but this
requires a larger number of choice variables resulting in a longer time to
solve.
As a compromise, we choose about 10 evenly spaced points on the curve leading
to an approximation with about $9$ line segments resulting in a worst-case
deviation of $4.15\%$ between the piece-wise linear approximation and the
actual function.
Since $w^2$ is a convex function, its piece-wise linear approximation can be
expressed using linear constraints with no integral variables.
However, since $-w^2$ is not a convex function, its piecewise linear
approximation requires us to utilize new binary integral variables to choose
the appropriate line segment, resulting in a MIP (Mixed Integer Program)
instead of an LP (Linear Program).
For a more detailed treatment of the techniques that we have used to remove the
two types of nonlinearities, the reader is referred to
\cite{williams1999model}.

%% file: Implementation.tex
\section{Model implementation and validation}

In this section, we discuss how our model outputs could be instantiated in a
real MapReduce framework using Hadoop as a representative MapReduce
implementation.
We have modified Hadoop in order to understand the accuracy of our model
outputs as well as to evaluate the benefits of our optimization.
Next, we discuss the implementation details of our changes made to Hadoop,
followed by validation results for our model based on this implementation.

\subsection{Implementing an execution plan in Hadoop}
\label{subsec:impl}

Recall from Section~\ref{sec:model} that in our model, a valid MapReduce
execution plan is defined as a set of data fractions \{$x_{ij}\}$ over all
links $(i,j)$ in the MapReduce cluster, where these \{$x_{ij}\}$ values satisfy
Equations~\ref{eq:frac1}, \ref{eq:frac2}, and \ref{eq:onekeyrule}.
To enable a valid execution plan to be instantiated in Hadoop, we make three
primary modifications to Hadoop:
(i) enforcing a tight coupling between data placement and task execution, so
that the work carried out on a node strictly depends on the fraction of the
data it receives as part of the execution plan;
(ii) controlling the push-phase data placement to implement the fraction of
data to be sent between each source-mapper pair;
and (iii) controlling the shuffle-phase data placement to send the fraction of
data between each mapper-reducer pair as per the execution plan.
We apply these changes to Hadoop version 1.0.1.
Here we discuss each of these changes in detail.
In addition, we also discuss how we achieve different barrier configurations
(global, local and pipelining) as part of the Hadoop framework.

\subsubsection{Coupling data placement and task execution}

Our model assumes that the data placement in the push and shuffle phases
uniquely determines the task execution in the map and reduce phases
respectively.
In the map phase, Hadoop attempts to satisfy this assumption using the
so-called ``locality optimization''~\cite{mapreduce}, whereby a map task is
scheduled on a mapper node that already hosts the data for that task.  
However, this optimization is not strictly enforced in Hadoop.
For example, if a mapper node (a \verb=TaskTracker= in Hadoop parlance) has
already completed map tasks for all of the input data that it hosts, it may be
assigned map tasks that read inputs from remote nodes.
In order to isolate the effects of our optimized plans from such dynamic
mechanisms, we add a configuration option (which we call \verb=LocalOnly= for
brevity) to Hadoop to disable assignment of remote tasks.
This involves simple modifications to the \verb=JobInProgress= class, which is
responsible for maintaining information about the map and reduce tasks
associated with a running job, and for responding to the scheduler's request
for new map and reduce tasks.
The existing implementation already supports the data-locality optimization.
Specifically, each map task in Hadoop reads its input from an
\verb=InputSplit=, which exposes a \verb=getLocations= method to the scheduler.
This method reports the host(s) on which the task can be considered local; for
an \verb=InputSplit= backed by a file block in the Hadoop Distributed File
System (HDFS) for example, \verb=getLocations= would return the locations of
the replicas of that block.
Additionally Hadoop allows users to specify the topology of the network.
Hadoop uses these two mechanisms to estimate the distance between a
\verb=TaskTracker= and an \verb=InputSplit=.
Our implementation simply checks the status of the \verb=LocalOnly=
configuration parameter, and if it is set, forces the \verb=JobInProgress= to
return to the scheduler only those tasks that are completely local to the
requesting \verb=TaskTracker=.

For reduce tasks, on the other hand, there is no data-locality optimization, as
shuffle communication follows an all-to-all pattern in the general case due to
the one-reducer-per-key requirement.
We couple the intermediate data placement to reduce task execution by
establishing a mapping between \verb=TaskTracker=s and reduce tasks.
For ease of implementation, we encode this mapping using Hadoop's
\verb=Configuration= API, recording which reduce partitions each
\verb=TaskTracker= is allowed to run.
Then when the scheduler requests a new reduce task, \verb=JobInProgress=
returns only tasks in this set.
If no mapping is specified for a \verb=TaskTracker=, then reduce task
assignment behaves as it does in the default Hadoop.

An additional case where Hadoop might break our assumption occurs when it
launches speculative (i.e., backup) tasks.
Hadoop already provides a configuration option to disable such speculative
tasks, and we employ this control directly.

\subsubsection{Controlling data placement in the push phase}

Now that we have established the tight coupling between data placement and task
execution, we need a way to control the data placement according to the
$x_{ij}$ values specified by the execution plan.
For map tasks, we achieve this by implementing a custom \verb=InputFormat= and
a corresponding custom \verb=InputSplit=.
The \verb=getLocations= method behaves as mentioned earlier, by returning the
host name for the \verb=TaskTracker= on which that task should run.
The \verb=InputSplit= is also responsible for providing its map task with a
\verb=RecordReader= for reading inputs.
Our \verb=InputSplit= encodes a list of data sources along with the amount of
data to read from each, and it builds a \verb=RecordReader= that reads from
each of these data sources concurrently over TCP sockets.
These sockets feed a producer-consumer queue, and the map tasks can read from
this queue as a single stream.
For ease of implementation, our \verb=RecordReader= connects to our own simple
data source server using TCP sockets.

The traditional \verb=FileInputFormat= constructs \verb=InputSplit=s that each
closely follow HDFS block boundaries, requiring no user control.
Our \verb=InputFormat=, on the other hand, reads a user-provided push plan (as
produced by our optimization), and builds a set of \verb=InputSplit=s that
achieve the planned push distribution.
As an example, if the plan calls for a mapper $M_1$ to read 3/4 of its data
from data source $S_1$ and 1/4 from data source $S_2$, then the
\verb=InputSplit=s destined for $M_1$ will each read 3/4 of their data from
$S_1$ and 1/4 of their data from $S_2$.
The size of each individual \verb=InputSplit= is limited by a user-specified
parameter; in our experiments we limit the size to 64MB, the same size we use
for HDFS file blocks.

\subsubsection{Controlling data placement in the reduce phase}

To control the intermediate data placement for reduce tasks, we implement a
custom \verb=Partitioner= class that first partitions intermediate keys into
buckets in exactly the same way that a typical \verb=Partitioner= does (the
default simply uses a hash function).
We set the number of buckets significantly larger than the number of reduce
tasks, then assign an appropriate number of these small buckets to each reduce
task.
For example, if we have two reducers $R_1$ and $R_2$, and the plan calls for
$R_1$ to reduce 2/3 of the intermediate keys, then we assign 2/3 of these
buckets to the partition for $R_1$, and 1/3 of the buckets to the partition for
$R_2$.
This is possible because, as we discussed earlier, we establish a unique
mapping between partition numbers and \verb=TaskTrackers=.\footnote{This
approach assumes that the original user-provided partition function achieves
roughly equal-sized partitions.  This is true for many typical MapReduce
applications, particularly when the key space is large.}
We implement a convenience method to read a plan from a file, and use the
\verb=Configuration= API to configure the \verb=Partitioner= as well as the
partition-to-\verb=TaskTracker= mapping appropriately.

\subsubsection{Instantiating barrier configurations}
\label{subsec:hadoop_barriers}

Our model also allows us to instantiate different barrier configurations
(global, local and pipelining) at each phase boundary as part of the MapReduce
job execution.
We instantiate a subset of all possible barrier configurations in Hadoop:
Hadoop supports some of these barrier configurations by default, while we do
not consider some of the others that are either hard to implement within the
Hadoop framework, or are not immediately interesting.
We achieve the following barrier configurations:
\begin{myitemize} 
  \item
  Push/map barriers: To achieve a global barrier at the push/map phase
  boundary, we run a separate map-only job to enact the push, which uses the
  custom \verb=InputFormat= to read directly from the data sources, and writes
  directly to HDFS.
  Then we run the main job, which uses a regular \verb=FileInputFormat= to read
  directly from HDFS as a typical Hadoop MapReduce job would do.
  This is the same way the \verb=DistCP= tool from the Hadoop distribution
  copies files from one distributed file system to another.
  Pipelining is achieved by using the custom \verb=InputFormat= in the main job
  itself to read directly from the data sources to the mappers.
  We have not instantiated a local barrier at this phase boundary.

  \item
  Map/shuffle barriers: For global barriers, we set the Hadoop configuration
  parameter\newline{\tt mapred.reduce.slowstart.completed.maps} to 1.
  This parameter specifies the fraction of map tasks that must complete before
  any reduce tasks are scheduled, and is often used in shared clusters in order
  to keep reduce tasks from occupying scarce reduce slots while they are merely
  waiting for input data.
  By setting it to 1, we require that all mappers finish before any reducers
  start.
  Since the shuffle is actually carried out by reducers pulling their data, if
  no reducers start until the whole map phase finishes, then there is also no
  shuffle until the map phase finishes; i.e., map and shuffle phases are
  separated by a global barrier.
  Coarse-grained pipelining is essentially the default behavior of Hadoop as
  long as there are enough reduce slots to schedule all reduce tasks
  immediately.
  MapReduce Online~\cite{condie:nsdi2010} implements finer-grained pipelining
  between these phases, but we consider only Hadoop's coarse-grained pipelining
  here.
  We have not instantiated a local barrier at this phase boundary.

  \item
  Shuffle/reduce barriers: Here, local barrier is the default configuration of
  Hadoop, as a reducer can start as soon as it has finished receiving all of
  its input data without waiting for other reducers to finish the shuffle
  phase.
  Note that implementing pipelining at this boundary is difficult in general:
  the MapReduce programming model requires that each invocation of the reduce
  function be provided an intermediate key and \emph{all} intermediate values
  for that key.
  Relaxing the barrier at this phase therefore requires changes at the
  application level, as Verma et al.~\cite{verma2010} describe in detail.
  We do not implement the global barrier  at this phase boundary.
\end{myitemize}

\subsection{Model estimation and validation} 
\label{subsec:validate}

In this section, we show how the various parameters of our model can be
estimated from actual measurements.
Further, we validate our model by correlating the model predictions for
makespan of different execution plans with the actual measured makespan
achieved by our modified Hadoop implementation executing the corresponding
plans.

We use PlanetLab~\cite{planetlab} for our measurements, since it is a globally
distributed testbed and is representative of the highly distributed environment
that we are considering in this paper.
We measure bandwidth and compute capacities for a set of PlanetLab nodes, and
estimate the model parameters for our distributed experimental platform as
follows.
For each $(i,j) \in E$, bandwidth $B_{ij}$ is estimated by transferring data
over that link and measuring the achieved bandwidth.
For stable estimates, we used downloads of size at least 64\unit{MB} or a
transfer time of at least 60 seconds, whichever occurs first.
For each compute node $i$, we estimate its compute rate $C_i$ by measuring the
runtime of a simple computation over a list within a Python program, yielding a
rate of computation in megabytes per second.
This value is not directly useful on its own, but we use it to estimate the
{\em relative} speeds of different nodes which may be used as part of a
distributed MapReduce cluster.

We base these measurements on a set of eight physical PlanetLab nodes
distributed across eight sites, including four in the United States, two in
Europe, and two in Asia.
The unscaled $C_i$ values on these nodes range from as low as 9\unit{MBps} to
as high as about 90\unit{MBps}.
Table~\ref{tbl:plab-nodes} shows the intra-continental and inter-continental
link bandwidths for these nodes and highlights the heterogeneity that
characterizes such a highly distributed network.

\begin{table}
  \caption{Measured bandwidth (\unit{KBps}) of the slowest/fastest links between
  clusters in each continent.}\label{tbl:plab-nodes}
  \centering
  \begin{tabular}{| l | c | c | c | }
    \hline
    & US & EU & Asia \\
    \hline
    % spacing looks bad if it is just num/num
    US & 216~/~9405 & 110~/~2267 & 61~/~3305 \\ \hline
    EU & 794~/~2734 & 4475~/~11053 & 1502~/~1593 \\ \hline
    Asia & 401~/~3610 & 290~/~1071 & 23762~/~23875 \\ \hline
  \end{tabular}
\end{table}

To study the predictive power of this model for a range of the application
parameter $\alpha$, we implement a synthetic Hadoop MapReduce job that allows
direct control over this parameter.
Mappers in this job read a key-value pair and emit that same key-value pair an
appropriate number of times to achieve the user-specified $\alpha$ value.
For example, if $\alpha=0.5$, then this synthetic mapper would directly emit
only every other input key-value pair; with $\alpha=2$, it would emit every
input key-value pair twice.
This job uses an identity reducer.
This synthetic application also allows us to emulate computation heterogeneity
by carrying out a different amount of computation on each node based on a
user-provided parameter.

Using this synthetic job, we can study the predictive power of our model by
correlating model predictions and actual measured makespan using the following
approach.
Given the scaled $C_i$ values along with the bandwidth measurements from
PlanetLab, we can derive the model's prediction for makespan by using the model
equations (Equations~\ref{eq:frac1} to \ref{eq:makespan}, with substitutions
as needed for local barriers or pipelining).
We then run a MapReduce job using our prototype implementation in Hadoop.
We run this MapReduce job on a local cluster of eight nodes, where each node
has two dual-core 2800\unit{MHz} AMD Opteron 2200 Processors, 4\unit{GB} RAM, a
250\unit{GB} disk, and Linux kernel version 2.6.32.
The nodes are connected via Gigabit Ethernet and we use {\tt tc} to emulate the
network bandwidths measured on PlanetLab.
We use an emulated local testbed to achieve stable and repeatable experiments,
as well as due to the daily bandwidth limits and severe memory constraints on
PlanetLab, which prevent such data-intensive experiments.
Overall, we emulate the full heterogeneity of the PlanetLab environment by
using {\tt tc} to emulate network heterogeneity and our synthetic MapReduce
job to emulate computation heterogeneity as well as the parameter $\alpha$.

For our validation, we consider $\alpha$ values of 0.1, 1, and 2.
We consider two levels of network heterogeneity: the network conditions
measured from PlanetLab as well as no emulation (raw LAN bandwidths).
We also consider two levels of computation heterogeneity: those measured from
PlanetLab, as well as no emulation.
We consider four barrier configurations: G-P-L, P-P-L, P-G-L, and G-G-L, where
G, L, and P correspond to global barrier, local barrier and pipelining
respectively, and each barrier sequence (e.g., G-P-L) corresponds to the
sequence of barriers enacted at push/map, map/shuffle, and shuffle/reduce phase
boundaries respectively.
We include two types of execution plans: a uniform plan that distributes the
input and intermediate data uniformly among the mappers and reducers
respectively, as well as an optimized plan generated by our optimization
algorithm.
All of our plans use 256\unit{MB} of input data from each of the eight data
sources.
We use two mappers and one reducer on each of eight compute nodes.
Actual makespan varies from 175 seconds to 2849 seconds.

\begin{figure}%[thbp]
  \centering
  \includegraphics[width=1.0\columnwidth]{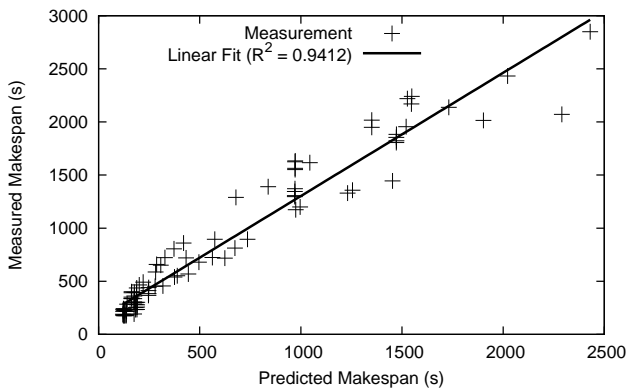}
  \caption{Measured makespan from the Hadoop MapReduce implementation on the
  local emulated testbed correlates strongly with the model-predicted
  makespan.\label{fig:validation}}
\end{figure}

The results of the validation are shown in Figure~\ref{fig:validation}.
This figure shows a strong correlation ($R^2$ value of 0.9412) between
predicted makespan and measured makespan.
In addition, the linear fit to the data points has a slope of 1.1464, which
shows there is also a strong correspondence between the absolute values of the
predicted and measured makespans.

%% file: OptEval.tex
\section{Benefits of optimized execution}

We provide an intuitive understanding of the optimization algorithm proposed in
Section~\ref{subsec:optimize} and evaluate the benefit of the execution plans
that it produces.
The two primary aspects that our optimization controls are how much data is
sent over which links in each of the push and shuffle phases.
Since task scheduling is tightly coupled to data placement, once the
push and shuffle phases are specified the entire execution plan is determined.

Our proposed optimization has two key properties.
First, it minimizes the {\em end-to-end} makespan of the whole MapReduce job,
which includes the time from the initial data push to the final reducer
execution.
Thus, its decisions may be suboptimal for individual phases, but will be
optimal for the overall execution of the MapReduce job.
Second, our optimization is {\em multi-phase} in that it controls the data
dissemination across both the push and shuffle phases; i.e., it outputs the
best way to perform both the push and shuffle phases so as to minimize
makespan.
To understand the benefits of our end-to-end multi-phase optimization relative
to other schemes, we answer the following questions.
\begin{myitemize}
  \item How beneficial is it to optimize end-to-end makespan as opposed to {\em
    myopically} minimizing the time for a single phase?
    (Section~\ref{subsec:e2e})

  \item How beneficial is it for our optimization to control multiple
    phases---i.e., both the push and shuffle phases---as opposed to just a {\em
    single phase}? (Section~\ref{subsec:multiphase})

  \item How much better are optimal execution plans compared with using
    uniform data placement with no explicit optimization?
    (Sections~\ref{subsec:e2e} and \ref{subsec:multiphase})

  \item What is the impact on optimized makespan of relaxing the barriers
    between phases?  (Section~\ref{subsec:barriers})

  \item What is the impact of the distributed network environment on the
    optimized execution plan? (Section~\ref{subsec:network})

  \item How does our optimized execution plan compare to Hadoop?
    (Section~\ref{subsec:hadoop})
\end{myitemize}
 
\subsection{Experimental setup}
\label{subsec:exptsetup}

To evaluate our optimization schemes, we use PlanetLab measurements to create
the different distributed network environments that we use in our experiments.
Our network environments vary from a single data center that is relatively
homogeneous, to a diverse environment comprising eight globally distributed
data centers.
To realistically model these environments, we use actual measurements of the
compute speed and link bandwidths from  PlanetLab nodes distributed around the
world, including four in the US, two in Europe, and two in Japan with compute
rates and interconnection bandwidths as described in
Section~\ref{subsec:validate}.
Using these measurements, we generate the following specific network
environments:
\begin{myitemize}
  \item
  {\em Local data center:} This setup consists of one local cluster with eight
  nodes of each type (source, mapper, reducer), and corresponds to the
  traditional local MapReduce execution scenario.
  This cluster is based purely on nodes in the US, specifically at tamu.edu.

  \item
  {\em Intra-continental data centers:} This setup consists of two data centers
  located within a continent---all nodes are in the US at tamu.edu and
  ucsb.edu.
  This setup corresponds to a more distributed topology than the local cluster
  scenario.

  \item
  {\em Globally distributed data centers:} In this setup, nodes span the entire
  globe (California, Texas, Germany, Japan), introducing much greater
  heterogeneity and wide-area network bandwidths and latencies.
  We considered two different global environments: one with four data centers
  (at ucsb.edu in California, tamu.edu in Texas, tkn.tu-berlin.de in Germany,
  and pnl.nitech.ac.jp in Japan), and another with eight data centers (same as
  earlier plus hpl.hp.com in California, uiuc.edu in Illinois, essex.ac.uk in
  the UK, and wide.ad.jp in Japan).
  This allows us to compare the impact of scaling up the number of locations.
\end{myitemize}

For each of the above setups, the total number of nodes is held constant at
eight.
In some cases, where we did not have sufficiently many nodes to meet this
requirement (e.g., for the local data center setup), we added replica nodes to
the setup with the measured node/link characteristics of the corresponding real
nodes.
In addition, we held the number of data sources fixed, allocating these data
sources to clusters in the same proportion as mappers and reducers.
The total amount of input data per data source was held constant throughout.

The globally distributed environment with eight data centers that is described
above is most appropriate for studying the general properties of our
optimization, as it closely resembles the highly-distributed settings that we
focus on in our work.
Consequently, we use this environment in all our experiments, except in
Section~\ref{subsec:network} where we study the impact of distribution of
network resources on our optimization.
Therefore, in that section, we use all of the above environments, from the
least distributed single data center to the most distributed eight data center
environment.
  
\subsection{End-to-End versus myopic}
\label{subsec:e2e}

\begin{figure*}[htbp]
  \centering
  \subfigure[$\alpha$=0.1]{
    \includegraphics[width=0.65\columnwidth]{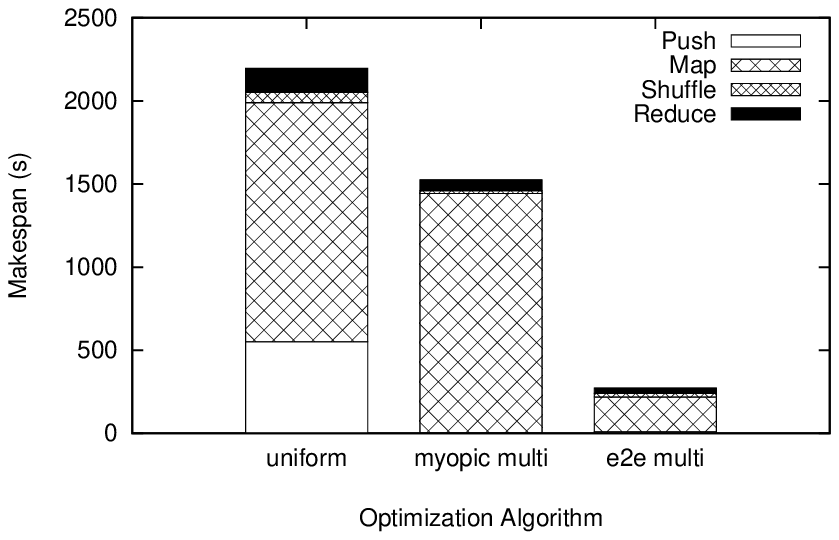}
  }
  \subfigure[$\alpha$=1]{
    \includegraphics[width=0.65\columnwidth]{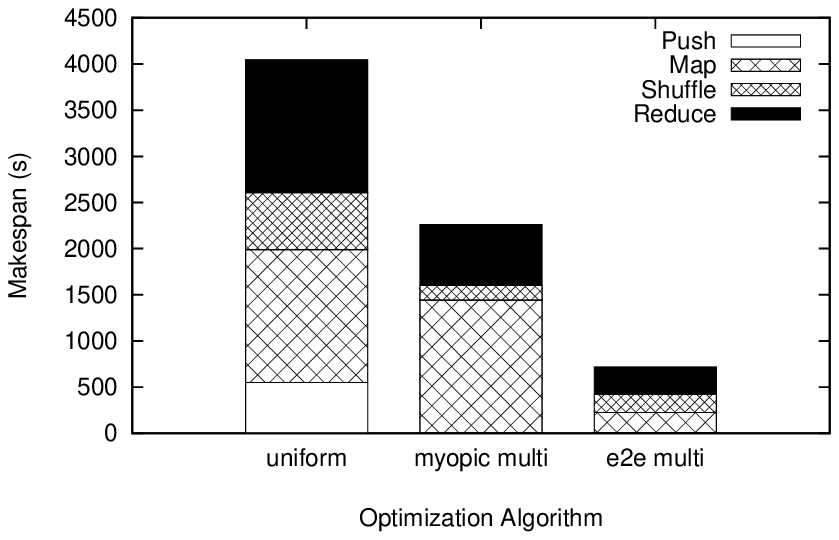}
  }
  \subfigure[$\alpha$=10]{
    \includegraphics[width=0.65\columnwidth]{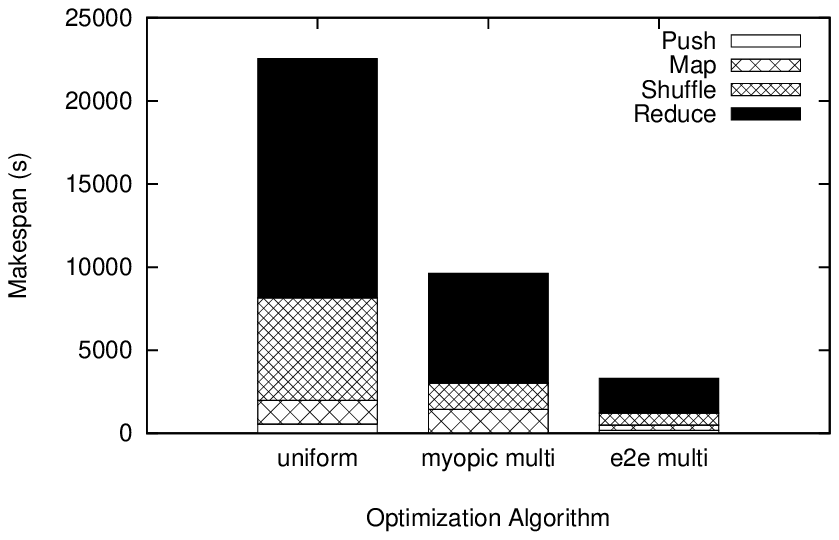}
  }
  \caption{Our proposed end-to-end multi-phase optimization (e2e multi)
  outperforms the uniform scheme by a large margin (82-87\%), while myopic
  multi-phase optimization (myopic multi) outperforms the uniform scheme by a
  smaller but still significant margin (30-57\%).
  Our end-to-end multi-phase scheme outperforms myopic multi-phase by a margin
  of 65-82\%, demonstrating the value of end-to-end optimization.}
 \label{fig:benefit_of_e2e}
\end{figure*}

The distinction between end-to-end and myopic optimization is the {\it
objective function} that is being minimized.
With end-to-end, the objective function is the overall makespan of the
MapReduce job, whereas a myopic optimizer uses the time for single data
dissemination phase (push or shuffle) as its objective function.
Myopic optimization is a localized optimization, e.g., pushing data from data
sources to mappers in a manner that minimizes the data push time.
Such local optimization might result in suboptimal global execution times by
creating bottlenecks in other phases of execution.
Note that myopic optimization can be applied to both push and shuffle phases in
succession.
That is, the push phase is first optimized to minimize push time and then the
shuffle phase is optimized to minimize shuffle time assuming that the input
data has been pushed to mappers according to the first optimization.

The myopic optimizations described above can be achieved by modifying our
formulation in Section~\ref{sec:model} as follows.
Instead of minimizing makespan, we replace Equation~\ref{eq:makespan} with
alternate objective functions.
For optimizing the push time we use: minimize $\max_{j \in M}{\tpushend_{j}}$,
and to optimize the shuffle time we use: minimize $\max_{k \in
R}{\tshuffleend_{k}}$, where $M$ and $R$ are the sets of mapper and reducer
nodes respectively.
(Note that computing a myopic multi-phase plan requires solving several
optimizations in sequence.)

As a comparative baseline, we would also like to evaluate the makespan produced
by the uniform schedule that involves no optimization at all.
In a uniform schedule, we distribute the input data across the mappers
uniformly, and also distribute the intermediate data across the reducers
uniformly.
This could be expressed in our model using the following additional constraints:
\begin{eqnarray} 
  \mathrm{Uniform\ Push:} &  \forall i \in S, j \in M : x_{ij} = \frac{1}{|M|}\label{eq:uniform_push} \label{eq:uniformpush}\\
  \mathrm{Uniform\ Shuffle:} &  \forall j \in M, k \in R : x_{jk} = \frac{1}{|R|} \label{eq:uniform_shuffle} \label{eq:uniformshuffle}
\end{eqnarray}

The above constraints implicitly assume that the nodes and communication links
are homogeneous, so that the sources uniformly distribute their data to the
mappers, and the mappers uniformly distribute their data to the reducers.

In Figure~\ref{fig:benefit_of_e2e}, we show the makespan achieved in three
different cases: (i) for a uniform data placement; (ii) for a myopic,
multi-phase optimization, where the push and shuffle phases are optimized
myopically in succession; and (iii) for our end-to-end, multi-phase optimization
that minimizes the total job makespan.
Note that since both (ii) and (iii) are multi-phase, the primary difference
between them is that one is myopic and the other is end-to-end, helping us
determine the relative merits of end-to-end versus myopic approaches.
We evaluated these three schemes for different assumptions for $\alpha$.
We see that, for each $\alpha$, the myopic optimization reduces the makespan
over the uniform data placement approach (by 30, 44, and 57\% for $\alpha$ =
0.1, 1, and 10 respectively), but is significantly outperformed by the
end-to-end optimization (which reduces makespan by 87, 82, and 85\%).
This is because, although the myopic approach makes locally optimal decisions
at each phase, these decisions may be globally suboptimal, while our end-to-end
optimization makes globally optimal decisions.
As an example, for $\alpha$=0.1, while both the myopic and end-to-end
approaches dramatically reduce the push time over the uniform approach (by 99.4
and 98.5\% respectively), the end-to-end approach is also able to reduce the
map time substantially (by 85\%) whereas the myopic approach makes no
improvement to the map time.
A similar trend is evident for $\alpha$=10, where the end-to-end approach is
able to lower the reduce time significantly (by 68\%) over the myopic approach.
These results show the benefit of an end-to-end, globally optimal approach over
a myopic, locally optimal but globally suboptimal approach.

\subsection {Single-phase versus multi-phase}
\label{subsec:multiphase}
The distinction between single-phase and multi-phase is which phase (push,
shuffle, or both) is {\em controlled} by the optimization, and is orthogonal to
the end-to-end versus myopic distinction.
A single-phase optimization controls the data distribution of one phase---e.g.,
the push phase---alone, while using a uniform data distribution for the other
communication phase.
A single-phase optimization is myopic if it minimizes the time for that phase
alone.
However, it could also be end-to-end if it optimizes the phase so as to achieve
the minimum overall makespan.
A single-phase optimization may be achieved in our model by using one of the
uniform push or shuffle constraints (Equation~\ref{eq:uniformpush} or
\ref{eq:uniformshuffle}) to constrain the data placement for one of the phases,
while allowing the other phase to be optimized.

\begin{figure*}[htbp]
  \centering
  \subfigure[$\alpha$=0.1]{
    \includegraphics[width=0.65\columnwidth]{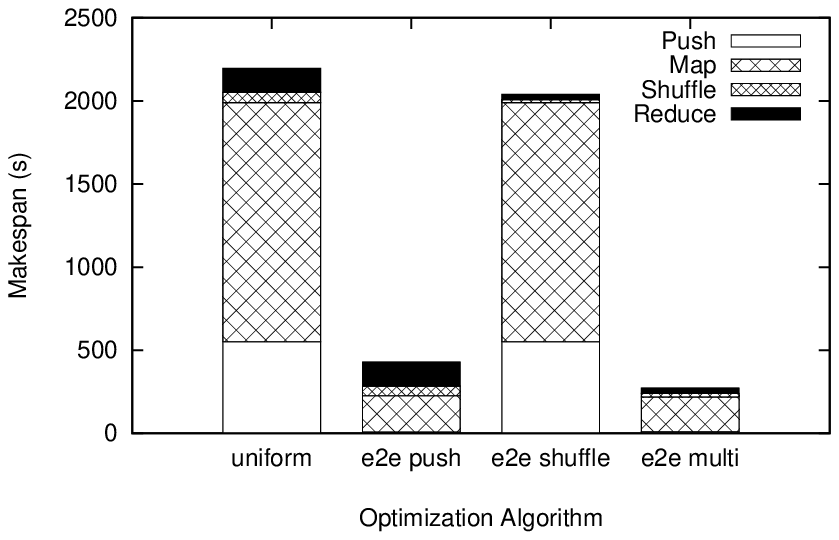}
  }
  \subfigure[$\alpha$=1]{
    \includegraphics[width=0.65\columnwidth]{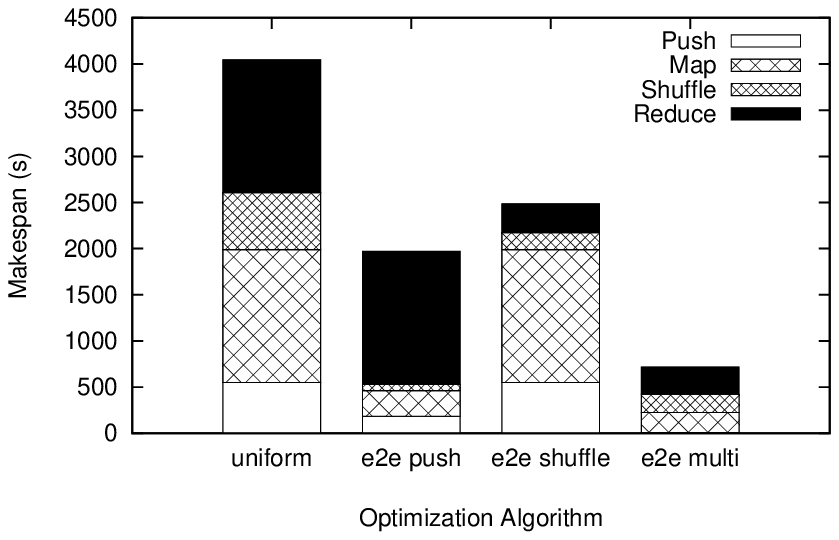}
  }
  \subfigure[$\alpha$=10]{
    \includegraphics[width=0.65\columnwidth]{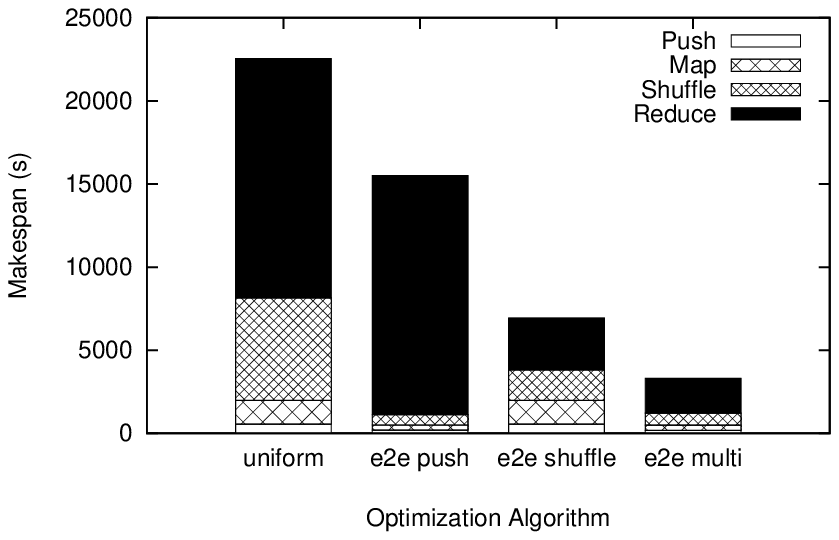}
  }
  \caption{Our end-to-end multi-phase optimization (e2e multi) performs better
  than end-to-end push (e2e push) and end-to-end shuffle (e2e shuffle),
  demonstrating the value of multi-phase optimization.
  In addition, all forms of optimization perform better than uniform.}
  \label{fig:benefit_of_multiphase}
\end{figure*}

In Figure~\ref{fig:benefit_of_multiphase}, we compare (i) a uniform data
placement, (ii) an end-to-end single-phase push optimization that assumes a
uniform shuffle,  (iii) an  end-to-end single-phase shuffle optimization that
assumes a uniform push, and (iv) our end-to-end multi-phase optimization.
Note that both the single-phase optimizations here are end-to-end optimizations
in that they attempt to minimize the total makespan of the MapReduce job.
The primary difference between (ii) and (iii) on the one hand and (iv) on the
other hand is that the former are single-phase and the latter multi-phase,
letting us evaluate the relative benefit of single- versus multi-phase
optimization.

In Figure~\ref{fig:benefit_of_multiphase}, we observe that across all $\alpha$
values, the multi-phase optimization outperforms the best single-phase
optimization (by 37, 64, and 52\% for $\alpha$=0.1, 1, and 10 respectively).
This shows the benefit of being able to control the data placement across
multiple phases.
Further, for each $\alpha$ value, optimizing the bottleneck phase brings greater
reduction in makespan than optimizing the non-bottleneck phase.
For instance, for $\alpha$=0.1, the push and map phases dominate the makespan
in the baseline (being about 25\% and 66\% of the total runtime for uniform,
respectively) and push optimization alone is able to reduce the makespan over
uniform by 80\% by lowering the runtime of these two phases.
On the other hand, for $\alpha$=10, the shuffle and reduce phases are dominant
(27\% and 64\% of total runtime for uniform, respectively) and optimizing these
phases via the shuffle optimization brings the makespan down by 69\% over
uniform.

An additional interesting observation from
Figures~\ref{fig:benefit_of_multiphase}(b) and (c) is that optimizing earlier
phases can have a beneficial impact on the performance of the later phases.
In particular, for $\alpha$=10, push optimization also brings down the shuffle
overhead (by 90\%), even though the push and map phases themselves have minimal
contribution to the makespan.
This is because the location of the mappers to which data is pushed has an
impact on how data is shuffled to reducers.
By influencing the data placement across multiple phases, our multi-phase
optimization is able to perform even better by optimizing both the bottleneck
as well as non-bottleneck phases.
In particular, when there is no prominent bottleneck phase ($\alpha$=1), the
multi-phase optimization outperforms the best single-phase optimization
substantially (by 64\%).
These results show that the multi-phase optimization is able to automatically
optimize the execution independent of the application characteristics.

%% file: ModelPreds.tex
\subsection{Relaxing barriers}
\label{subsec:barriers}

\begin{figure}[htbp]
  \centering
  \includegraphics[width=1.0\columnwidth]{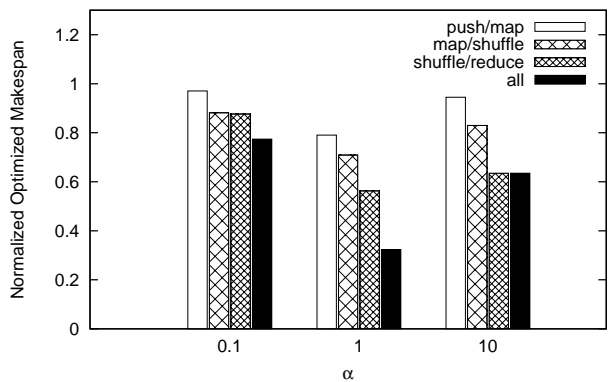}
  \caption{Predicted normalized makespan for optimized plans with various
  barrier configurations and different values of $\alpha$.
  All makespans are normalized relative to the optimal makespan for an
  all-global-barrier configuration.  Each bar represents the phase boundary at
  which the global barrier is relaxed to pipelining, with ``all'' corresponding
  to an all-pipelining configuration.
  \label{fig:benefit_of_pipelining}}
\end{figure}

Now we study the impact of relaxing barriers on the makespan predicted by our
model.
In particular, we focus on the impact of using pipelining vs. global barriers
at each phase boundary.
In Figure~\ref{fig:benefit_of_pipelining}, we show the normalized makespan for
a select set of different barrier configurations.
All predicted makespan values shown in the figure are normalized relative to
the optimal makespan derived for an all-global-barrier configuration, i.e., one
which has a global barrier at each phase boundary.
The bars in the figure show the effect of relaxing only a single global barrier
to pipelining at a time, at the push/map, the map/shuffle, and the
shuffle/reduce boundaries respectively, as well as  the all-pipelined
configuration, where all barriers are pipelined. We make two key observations.

\begin{myitemize}
  \item
  When phases are roughly ``balanced'' in terms of time taken, pipelining is
  most effective.
  This is because overlapping the execution of two balanced phases gives more
  opportunity for reducing their total execution time, compared to when one
  phase significantly dominates the other.
  For the parameters considered here, the phases are most closely balanced when
  $\alpha=1$, as can be seen from Figure~\ref{fig:benefit_of_multiphase}.
  Consequently, we see from Figure~\ref{fig:benefit_of_pipelining} that each
  barrier relaxation provides the greatest benefit when $\alpha=1$.

  \item
  Relaxing late-stage barriers---such as those between map and shuffle or
  between shuffle and reduce---is predicted to have a greater benefit than
  relaxing barriers between push and map stages.
  The reason is that our optimization is more constrained in data placement
  during the shuffle phase than the push phase due to the one-reducer-per-key
  constraint, and hence pipelining finds {\em more} opportunity to hide the
  latency of the shuffle phase with its adjoining computational phases (map or
  reduce).
  This phenomenon can also be observed from
  Figure~\ref{fig:benefit_of_multiphase}, where we see that the shuffle time is
  higher than the push time with our optimization (e2e multi), particularly for
  $\alpha$=1 and 10.
\end{myitemize}

To summarize, relaxing barriers is most useful when the execution phases are
roughly balanced, and for later phase boundaries, where there is more
opportunity for latency hiding over the optimal plan with global barriers.

\subsection{Distribution of network resources}
\label{subsec:network}
To evaluate the effect of distribution of network resources, we derive
optimized execution plans using our proposed (end-to-end multi-phase)
optimization algorithm for {\em all} of the network environments described in
Section~\ref{subsec:exptsetup}, starting from a relatively homogeneous
environment with a single data center, to an intra-continental setup consisting
of two data centers in the US, to globally-distributed setups with four or
eight data centers.
\begin{figure*}[htbp]
  \centering
  \subfigure[$\alpha$=0.1]{
    \includegraphics[width=0.65\columnwidth]{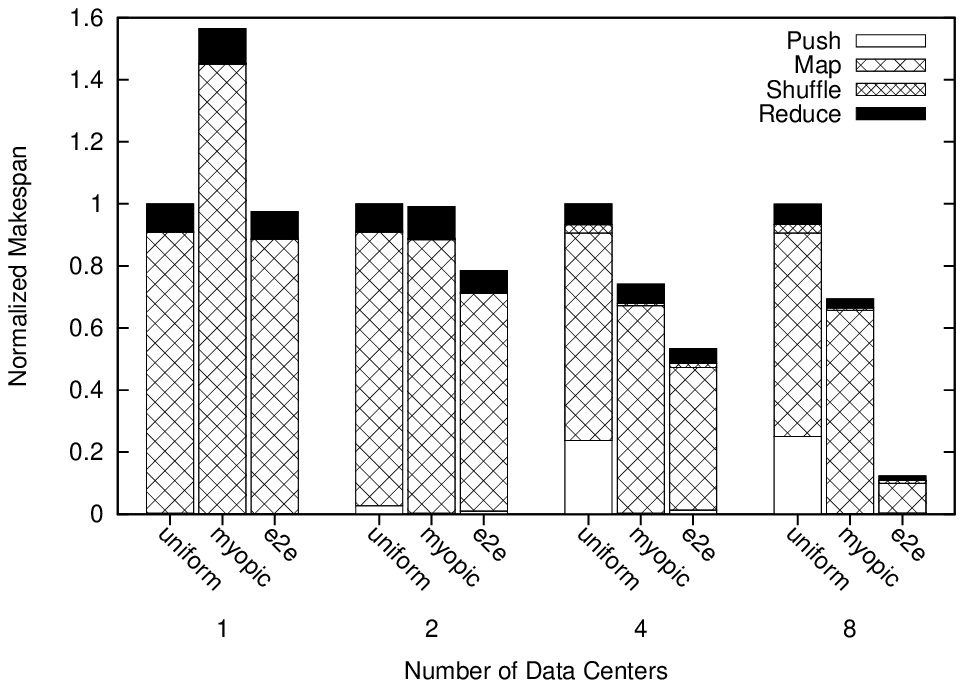}
  }
  \subfigure[$\alpha$=1]{
    \includegraphics[width=0.65\columnwidth]{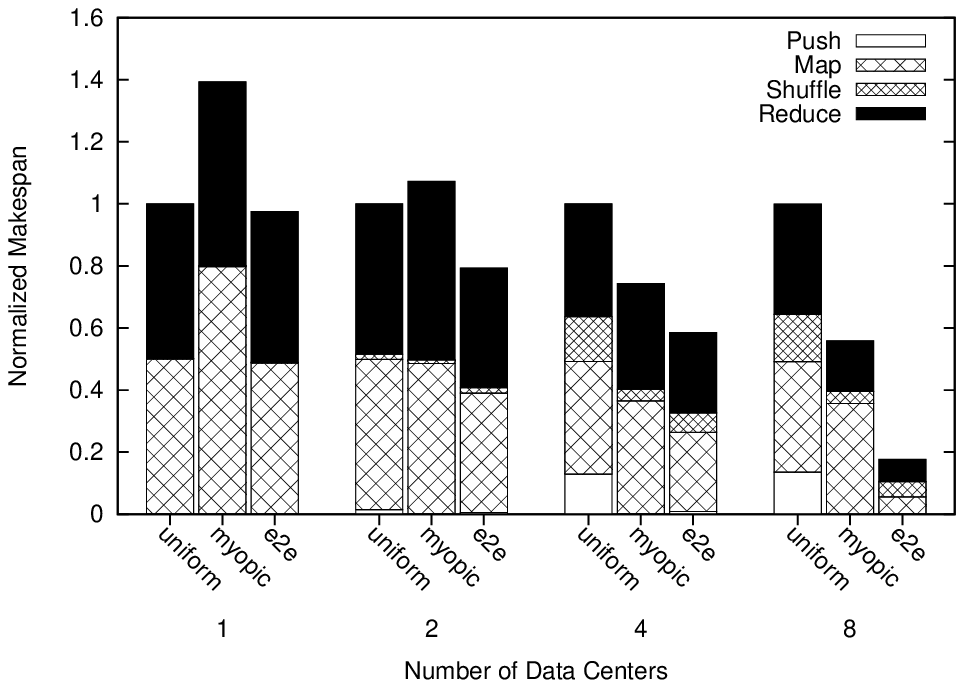}
  }
  \subfigure[$\alpha$=10]{
    \includegraphics[width=0.65\columnwidth]{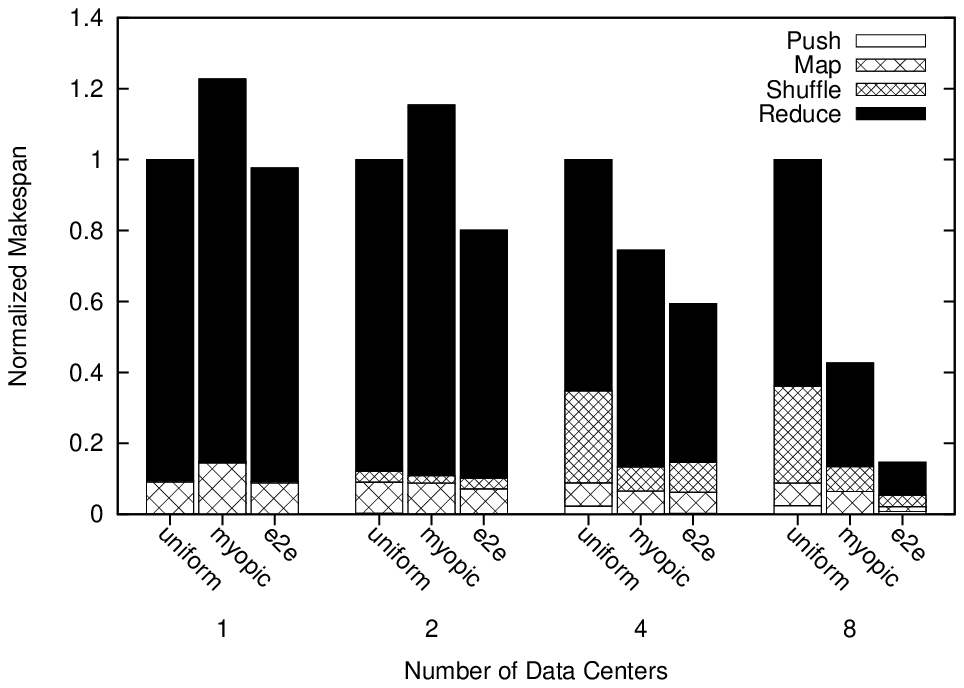}
  }
  \caption{A comparison of myopic and end-to-end optimization relative to the
  uniform baseline for different network environments and different values of
  $\alpha$.
  The makespans are normalized by computing the ratio of the actual makespan to
  the makespan of the corresponding uniform schedule.
  A global barrier is used between phases in all cases.}
  \label{fig:effect_of_environment}
\end{figure*}

Figure~\ref{fig:effect_of_environment} shows the makespan for the myopic
optimization that successively optimizes the execution times of the push phase
and the shuffle phase and the end-to-end optimization that optimizes makespan
by considering all phases at once.
Both the myopic and end-to-end optimizations are compared against a baseline
that performs uniform push and shuffle.

If the network environment is a single data center with relative homogeneity of
compute rates and network bandwidths, the uniform baseline does almost as well
as end-to-end optimization for all values of $\alpha$.
Note that this provides supporting evidence for why Hadoop's use of a largely
uniform schedule is quite effective in homogeneous environments.
Interestingly, some optimization is worse than none, as myopic optimization
does worse than uniform.
Intuitively, myopic optimization reacts to rectify small communication
imbalances, but this can in turn create larger computational imbalances among
the map and reduce tasks, resulting in a longer makespan.
This effect can be seen from the figure by the increased map and reduce times
for myopic for $\alpha$=0.1 and 10 respectively.

As the network environment becomes more diverse with more and more data
centers, both myopic and end-to-end optimizations start to perform better than
uniform, as uniform fails to account for the diversity of the environment.
As expected, end-to-end performs the best with makespans that are smaller by
82-87\% over uniform and by 65-82\% over myopic.

In summary, our results show that our optimization derives much greater
opportunity for improvement as the diversity and heterogeneity of the
environment increases, while reducing to a largely uniform placement for
tightly-coupled, homogeneous environments, where myopic optimization may
actually hurt performance.

\subsection{Comparing our optimization to Hadoop}
\label{subsec:hadoop}

We compare the results of our optimization against \textit{vanilla Hadoop},
which represents a typical unmodified Hadoop execution.

\subsubsection{Experimental setup}
For these experiments, we use the 8-node cluster with the emulated links
bandwidths of the distributed PlanetLab environment as described in
Section~\ref{subsec:validate}.
Each of our eight physical nodes hosts two map slots and one reduce slot, as
well as an HDFS datanode.
Each physical node also runs a simple TCP server to act as a data source.
We use our modified Hadoop implementation described in
Section~\ref{subsec:impl}, using largely default Hadoop configuration options,
aside from increasing \verb=io.sort.mb= to 200 and increasing the Java heap
size for worker processes to 800\unit{MB}.
We also set \newline\verb=dfs.replication= to 1 to prevent replication over
emulated slow links, which can have a pronounced adverse impact on performance
(See Section~\ref{subsec:repl}).
Since the emulated environment exhibits no inherent hierarchical structure,
there is no direct way to model it using Hadoop's rack-oriented model.
Therefore, for Hadoop's network topology configuration, we use the default that
models each node as being co-located in the same rack.

To focus our comparison on the efficacy of our optimized plans, we implement
the push phase for both vanilla Hadoop and our optimization using the same type
of \verb=InputSplit= as described in Section~\ref{subsec:impl}.
To provide vanilla Hadoop with a competitive baseline, we take advantage of
Hadoop's existing data-locality optimization, using the \verb=getLocations=
method of our \verb=InputSplit= to hint that Hadoop should push data from a
data source to the most local compute node.
In our testbed, since map tasks run on the same physical nodes as our data
sources, we hint that Hadoop should move data locally from the data source
server directly into the local HDFS data node.
This is logically identical to Hadoop's existing data-locality optimization,
but applied to the emulated wide-area setting.
In addition, we allow vanilla Hadoop to use its dynamic mechanisms such as
speculative task execution and non-local work stealing to avoid stragglers and
idle resources.

For our optimization, we provide our modified Hadoop implementation with the
exact plans produced by our optimization.
The optimized plan is derived using the model parameters for the underlying
platform and the application, and the model uses a G-P-L barrier configuration
at the consecutive phase boundaries to capture Hadoop's execution behavior.
In order to ensure that Hadoop strictly follows these plans, we turn on our
\verb=LocalOnly= configuration option (see Section~\ref{subsec:impl}).
Further, to understand the benefit of our offline execution plan, we turn off
Hadoop's dynamic mechanisms mentioned above for our optimization.
As a baseline, we also compare vanilla Hadoop and our optimization to a uniform
execution plan, which uniformly pushes and shuffles data to mappers and
reducers respectively.

\subsubsection{Applications}

We implement three MapReduce applications for this evaluation, which vary in
terms of their application characteristics, particularly, the expansion factor
$\alpha$:

\noindent{\em (1) Word Count.} This application takes as input a set of
documents and produces as output, for each term in the set, the number of
occurrences of that term.
Map tasks receive a plain text document and tokenize it, then count the number
of occurrences of each term.
For each term $t$, the mapper emits the key-value pair $(t, f(t))$ where $f(t)$
denotes the number of times term $t$ occurred.
We apply the in-mapper-combining pattern, described by Lin and
Dyer~\cite{lin2010ditp}.
Reducers receive key-value pairs of the form $(t, [f(t)])$ where $t$ is the
term and $[f(t)]$ denotes a list of all counts for that term.
The reducer simply sums up this list and emits as final output a tuple with $t$
as the key, and the sum of all counts as the value.
This application exhibits high aggregation; $\alpha=0.09$.
As inputs for this application, we use plain-text eBooks from Project
Gutenberg~\cite{gutenberg}.
The total input size is roughly 16.5\unit{GB}, spread across 48,000
eBooks.\footnote{Note that Project Gutenberg hosts fewer than 48,000 books at
the time of writing.  To reach this data size, we gather a large fraction of
the available books, and include each in our input data twice.}

\noindent{\em (2) Sessionization.} This application takes as input a collection
of Web server logs and produces as output, for each user, the sequence of
``sessions'' for that user.
At its core, this application is a large distributed sort.
Here the map function receives a single server log entry $v$, which it then
parses into a user identifier $id$ and timestamp $t$.
As intermediate data, mappers emit the composite key $(id, t)$ along with the
unchanged value $v$.
This application uses a custom \verb=SortComparator= to sort first by $id$,
then by $t$.
Intermediate key-value pairs are grouped using a custom
\verb=GroupingComparator= that groups only on the $id$; all log entries for a
single $id$ are then presented to a single call of the reduce function.
The system ensures that these values are delivered in sorted order by timestamp
$t$, and the reduce function simply determines the boundary between sessions
for that user by looking for sufficiently large gaps in this value.
There is no opportunity for aggregating (or expanding) intermediate results in
this application, as the mapper simply routes data to reducers.
Therefore $\alpha=1.0$.
We use a portion of the WorldCup98 trace~\cite{wc98} (roughly 5\unit{GB}
spanning 60 million log entries) as the input data for this application.

\noindent{\em (3) Full Inverted Index.} This application takes as input a
collection of documents, each represented as a pair of document identifier $id$
along with a sequence of word identifiers $w$.
It produces as output, for each word $w$, the complete list of documents in
which that word occurs, as well as the position within those documents.
The implementation is modeled after the example from Lin and
Dyer~\cite{lin2010ditp}.
This application also uses a custom \verb=SortingComparator= and
\verb=GroupingComparator= to rely as much as possible on the underlying
MapReduce system for its data movement.
This application expands the input data by adding additional information
regarding each term to the index, yielding $\alpha=1.88$.
As input, we use the same set of eBooks as for the Word Count application, but
preprocessed to remove stop words, and replace terms with an integer term
identifier; in essence a simple forward index.
This data again spans 48,000 books, but the more concise representation yields
a total input size of roughly 4\unit{GB}.

\subsubsection{Experimental results}

\begin{figure}[htbp]
  \centering
  \includegraphics[width=1.0\columnwidth]{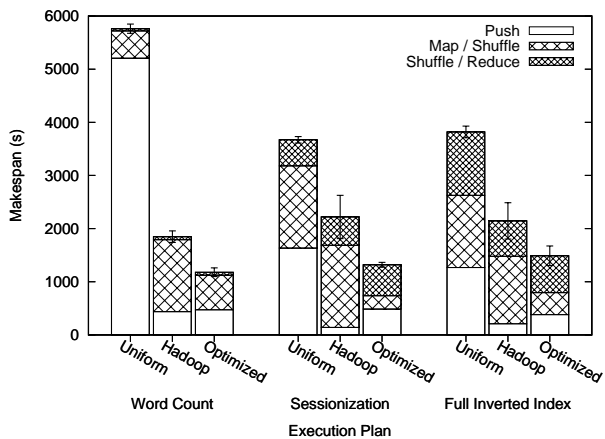}
  \caption{Actual makespan for three sample applications on our local testbed.
  Error bars reflect 95\% confidence intervals.}
  \label{fig:versus_hadoop}
\end{figure}

Figure~\ref{fig:versus_hadoop} shows the results of our comparison for the
three applications.
Note that for Hadoop, since the shuffle phase is partially overlapped with the
map and reduce phases, we depict only three phases in each bar in the graph:
the push phase, the overlapped map/shuffle phases, and the overlapped
shuffle/reduce phases.
We see that across all applications, while vanilla Hadoop substantially
outperforms the uniform execution plan (by 68, 40, and 44\% for Word Count,
Sessionization, and Full Inverted Index respectively), Hadoop executing our
optimized plan achieves a further improvement of 36, 41, and 31\% over vanilla
Hadoop for the same applications.
Further, we see how vanilla Hadoop makes myopic decisions.
Hadoop reduces the push time substantially (by 92, 91, and 83\% for Word Count,
Sessionization, and Full Inverted Index, respectively) over uniform.
However, our optimization, while increasing the push time over vanilla Hadoop,
achieves more significant reduction in end-to-end makespan than vanilla Hadoop
by better optimizing the map, shuffle, and reduce phases.
Thus, overall, we find that Hadoop executing our offline end-to-end,
multi-phase optimal execution plan outperforms vanilla Hadoop using its dynamic
mechanisms.

\subsubsection{Enhancing optimized plan dynamically} 

Our optimization provides an offline execution plan based on information about
the underlying infrastructure available before the job begins execution.
As mentioned above, Hadoop provides two dynamic mechanisms to modify the
initial execution plan based on the observed runtime behavior of the network and
nodes:  (i) {\em speculative task execution}, where another copy of a straggler
task is launched on a different node; and (ii) {\em work stealing}, where a
node may request a non-local task if that node is idle.
Here, we evaluate the benefit of using dynamic mechanisms in addition to the
static execution plan derived from the results of our optimization.

\begin{figure}[htbp]
  \centering
  \includegraphics[width=1.0\columnwidth]{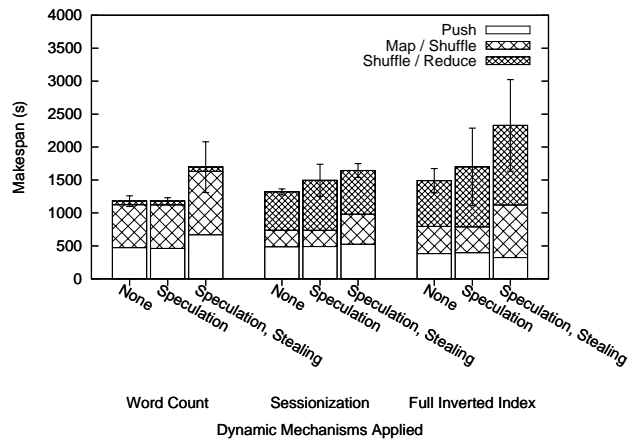}
  \caption{Effect of Hadoop's dynamic scheduling mechanisms applied atop our
  optimized static execution plan for three sample applications.
  Error bars reflect 95\% confidence intervals.
  \label{fig:effect_of_dynamic_mechanisms}}
\end{figure}

\begin{figure}[htbp]
  \centering
  \includegraphics[width=1.0\columnwidth]{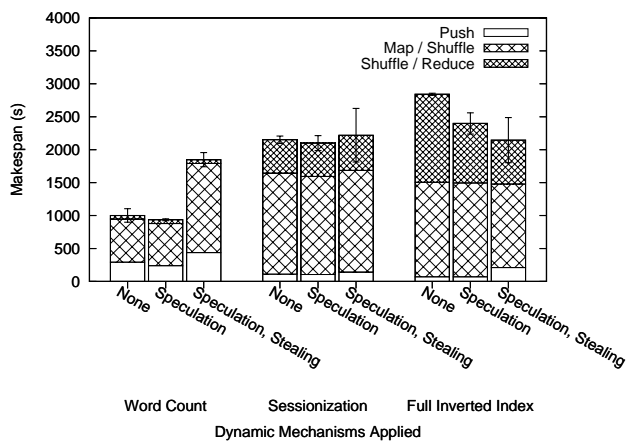}
  \caption{Effect of Hadoop's dynamic scheduling mechanisms applied atop a
  competitive Hadoop baseline plan, one that pushes from each data source to
  the most local mapper, and distributes intermediate keys uniformly across all
  reducers.
  Error bars reflect 95\% confidence intervals.
  \label{fig:effect_of_dynamic_mechanisms_atop_vanilla_hadoop}}
\end{figure}

Figure~\ref{fig:effect_of_dynamic_mechanisms} shows the impact of enabling
these two dynamic mechanism on the performance of the optimized plan for each
of the three sample applications.
Additionally, Figure~\ref{fig:effect_of_dynamic_mechanisms_atop_vanilla_hadoop}
shows the impact of enabling these mechanisms atop a competitive Hadoop
baseline plan.
We find from these two figures that, applied on its own, speculation does not
statistically significantly degrade performance in any case, while it
significantly improves performance in one case.
On the other hand, the addition of work stealing to speculation never
statistically significantly improves performance over speculation alone,
while in some cases---specifically Word Count---stealing significantly
degrades performance.
In fact, there is only one case where the combination of speculation and
stealing is statistically significantly better than the static baseline: Full
Inverted Index with the Hadoop baseline.
The reason in this case is that the static plan myopically optimizes the push
phase, adversely impacting the much more dominant shuffle and reduce phases.
By enabling speculation and optionally stealing, however, Hadoop is able to
bypass the bottleneck network links and compute nodes by moving data over
faster links and placing tasks on faster nodes.

Although the dynamic mechanisms are helpful in such a case, they can also
degrade performance in other cases.
For example, the combination of speculation and stealing statistically
significantly worsens performance for two of the three applications when
applied atop an optimized plan, and for one of the three applications when
applied atop a Hadoop baseline plan.
For the optimized plans, this occurs because dynamic changes to the offline plan
can actually undermine the optimization.
After all, if the computed plan is optimal, then barring any changes to the
underlying infrastructure, no dynamic change could improve performance.
For the Hadoop baseline, the degradation occurs for the Word Count application,
for which the runtime is dominated by the push and map phases.
For such an application, Hadoop's myopic optimization is actually quite
effective, and dynamically deviating from this plan can yield significantly
worse performance as we see here.

Though it is not shown directly in the figures, it is noteworthy that the best
Hadoop performance is never statistically significantly better than the
performance with the optimized plan.
Hadoop's best performance relative to the optimized plan occurs for the Word
Count application, for which Hadoop with speculation (but not stealing) yields
a lower mean makespan than the optimized plan, but not statistically
significantly so.

These results overall show the strength of the statically enforced optimized
plan.
At the same time, they show how dynamic mechanisms can improve performance when
the initial plan is far from optimal, as is the case for the Full Inverted
Index application with the Hadoop baseline.
Such a situation could also arise if network or node conditions were to change
significantly, for example due to network congestion or changes in background
CPU load.
Developing dynamic mechanisms that improve performance in such cases without
adversely affecting the performance in other cases is an interesting direction
for future work.

\subsubsection{Impact of data replication across slow links}
\label{subsec:repl}

\begin{figure}[htbp]
  \centering
  \includegraphics[width=1.0\columnwidth]{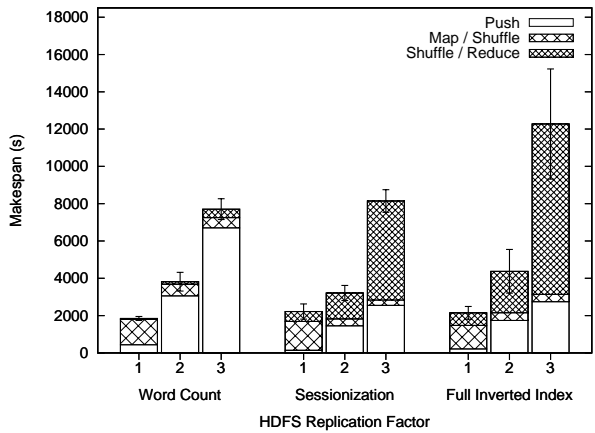}
  \caption{Effect of HDFS file replication for three sample applications.
  Error bars reflect 95\% confidence intervals.
  \label{fig:effect_of_replication}}
\end{figure}

In our model, we restrict replication to be intra-cluster, to avoid sending
redundant data across slow wide-area links.
Figure~\ref{fig:effect_of_replication} shows the impact of wide-area
replication on the performance of vanilla Hadoop for each of the three
applications.
As the figure shows, increasing replication substantially increases the cost of
data push, as well as the cost of the reduce phase due to the need to
materialize final results to the distributed file system.
While higher replication also yields a reduction in compute time in the map
phase due to greater scheduling flexibility, this improvement is dwarfed by the
increased communication costs.
However, replication across clusters would be useful for achieving higher fault
tolerance against geographically localized faults.
Such replication may also provide other opportunities for performance
optimization; e.g., work stealing may be directed to local tasks to avoid
high-overhead data communication.
Enhancing our model to incorporate cross-cluster replication, or complementing
our execution with other techniques such as checkpointing for intermediate
data~\cite{ko10} is an interesting direction for future work.

%% file: Related.tex
\section{Related Work}
\label{sec:relatedwork}

MapReduce implementations~\cite{mapreduce,hadoop} have traditionally been
deployed over a tightly-coupled cluster or data center, comprising largely
homogeneous compute resources connected over a local-area network.
Several research efforts have shown the impact on performance of MapReduce if
this assumption is broken.
Heterogeneity~\cite{heteromapred} in terms of node speeds was shown to have an
adverse impact on the default MapReduce scheduling performance.
Mantri~\cite{ananthanarayanan:osdi2010} further shows the impact of
heterogeneity in terms of machine and network characteristics as well as
application workload on the performance of MapReduce within a data center
environment.
Tarazu~\cite{ahmad2012} provides techniques for dynamic load balancing and
reducing network burstiness within  a tightly-coupled cluster of heterogeneous
compute nodes.
Our model is able to incorporate such heterogeneity in computation and
communication characteristics, not only within a single data center, but also
over a highly-distributed environment.
Further, some of these dynamic techniques can be applied at the data center
level, and are complementary to our model's outputs, which provide an initial
execution plan for a geographically distributed environment.

Our recent work~\cite{dmr11} has shown the performance impact on MapReduce in a
highly distributed environment, and explored multiple architectural choices for
deploying MapReduce based on network and application characteristics.
Hierarchical MapReduce~\cite{luo2011} adds a new ``Global Reduce'' stage to the
MapReduce semantics to aggregate results from a MapReduce job executed across
multiple clusters, though it avoids the issue of costly data push by assuming
data is small or already present in the local clusters.
In this paper, we present a more general analytic approach to explore some of
these tradeoffs, and our optimization provides the best execution plan based on
system characteristics.
Kim et al.~\cite{kim2011} consider Hadoop performance in an inter-cloud
environment, focusing on minimizing the end time of the shuffle phase, whereas
our end-to-end, multi-phase optimization focuses on minimizing the makespan of
the entire job.

Most existing work has examined the performance of MapReduce execution after
the push phase.
We explicitly model the push phase, which is particularly important due to the
presence of multiple data sources in our environment.
Most MapReduce implementations rely on a distributed file system, such as
GFS~\cite{gfs} or HDFS~\cite{shvachko2010hadoop}, from which mapper nodes can
pull their inputs.
The presence of a distributed file system also implicitly imposes a global
barrier between the push and map phases, where the mappers do not start
execution until all the input data has already been placed in the distributed
file system.
Hadoop effectively allows pipelining between the push and map phase and
coarse-grained pipelining between the map and shuffle phases (see
Subsection~\ref{subsec:hadoop_barriers}).
MapReduce Online~\cite{condie:nsdi2010} proposes finer-grained pipelining of
the map and shuffle phases, as well as pipelining between MapReduce jobs.
Verma et al.~\cite{verma2010} propose system changes to relax Hadoop's local
shuffle/reduce barrier.
This affects the semantics of the reduce phase, and they present techniques for
transforming applications to support this change.
Our model captures all these variations and enables us to compare the
performance of these choices across different phase boundaries.

Our model is data-oblivious; i.e., it does not assume knowledge of the input
data contents, but such knowledge could be exploited to further improve
performance.
SkewReduce~\cite{kwon:socc2010} presents the problem of application-specific
computational skew, where different parts of the input data may require
different amounts of computation resources.
Such data-dependent compute requirements can be incorporated in our model by
using data-dependent $C_i$ values.
CoHadoop~\cite{cohadoop} co-locates related data on the same node to improve
performance of certain applications.
Such an approach requires detailed knowledge of input data, which our model
does not assume.

Other work has focused on scheduling or fine-tuning MapReduce parameters to
provide better performance.
Sandholm et al.~\cite{sandholm:sigmetrics2009} present a dynamic priority-based
system for providing differentiated service to multiple MapReduce jobs.
Quincy~\cite{isard:sosp2009} is a framework for scheduling concurrent jobs to
achieve fairness while improving data locality.
Our focus is on optimizing the performance of individual job execution in a
more distributed environment.
Recent work~\cite{shivnath:socc} has proposed methods for automatically
fine-tuning Hadoop parameters to optimize job performance.
Our work takes a different approach, where we attempt to abstract away specific
implementation details, so that our model is general enough to capture the
abstraction behind many existing implementations.
Thus, our model is useful for comparing different design and architectural
choices, and some of its recommendations could be instantiated via some of the
existing work.

Elastisizer~\cite{herodotou2011} and STEAMEngine~\cite{steamengine} focus on
MapReduce provisioning within a cloud environment.
While Elastisizer uses offline profiling along with black-box and white-box
models to select the right cluster size, STEAMEngine uses both offline and
online job profiling along with dynamic scaling to provision and place the
jobs.
The focus of these works is on largely homogeneous cluster environments, as
opposed to multi-cluster environments like ours.
Further, provisioning is a complementary problem to the problem of performance
optimization within a given computation environment, addressed in this paper.

Fault tolerance in Hadoop has been addressed by increasing the availability of
intermediate data~\cite{ko10}.
MOON~\cite{moon:hpdc2010} explored MapReduce performance in a local-area
volunteer computing environment and extended Hadoop to provide improved
performance under low reliability conditions.
While our focus in this paper is on performance, achieving fault tolerance over
a highly distributed environment is an interesting area of future work.

%% file: Conc.tex
\section{Concluding Remarks}
\label{sec:conclude}

In this paper, we addressed the problem of executing MapReduce in a highly
distributed environment, comprising distributed data sources and computational
resources.
We developed a modeling framework to capture MapReduce execution in such a
distributed environment.
This framework is flexible enough to capture several design choices and
performance optimizations for MapReduce execution.
We proposed a model-driven optimization that has two key features: (i) it is
end-to-end as opposed to myopic optimizations that may only make locally
optimal but globally suboptimal decisions, and (ii) it can control multiple
MapReduce phases to achieve low runtime, as opposed to single-phase
optimizations that may control only individual phases.
Our model results showed that our optimization can provide nearly 82\% and 64\%
reduction in execution time over myopic and the best single-phase
optimizations, respectively.
We modified Hadoop to implement our model outputs, and using three different
MapReduce applications over an 8-node emulated PlanetLab testbed, we showed
that our optimized Hadoop execution plan achieves 31-41\% reduction in runtime
over a vanilla Hadoop execution.
Our model-driven optimization also provided several insights into the choice of
techniques and execution parameters based on application and platform
characteristics.
For instance, we found that an application's data expansion factor $\alpha$ can
influence the optimal execution plan significantly, both in terms of which
phases of execution are impacted more and where pipelining is more helpful.
Our results also showed that as the network becomes more distributed and
heterogeneous, our optimization provides higher benefits (82\% for globally
distributed sites vs. 37\% for a single local cluster over myopic
optimization), as it can exploit heterogeneity opportunities better.